\documentclass[12pt]{article}
\usepackage{amsmath}
\usepackage{amsfonts}
\usepackage{amssymb}
\usepackage{graphicx}

\def\ffrac#1#2{\textstyle{#1\over#2}\displaystyle}

\begin{document}
\setcounter{page}{0} \topmargin 0pt
\renewcommand{\thefootnote}{\arabic{footnote}}
\newpage
\setcounter{page}{0}

\begin{titlepage}

\begin{center}
{\Large {\bf Unusual Corrections to Scaling in Entanglement Entropy}}\\

\vspace{2cm}
{\large John Cardy$^{a,b}$\\
Pasquale Calabrese$^{c}$\\}  \vspace{0.5cm} {\em $^{a}$Rudolf
Peierls Centre for Theoretical Physics\\ 1 Keble Road, Oxford OX1
3NP, UK\\}  \vspace{0.2cm} {\em $^{b}$All Souls College, Oxford\\}
\vspace{0.3cm} {\em $^{c}$Dipartimento di Fisica, Universit\`a
di Pisa, and INFN, 56127 Pisa, Italy}

\vspace{2cm}

February 2010

\end{center}

\vspace{1cm}

\begin{abstract}
\noindent We present a general theory of the corrections to the
asymptotic behaviour of the R\'enyi entropies
$S_A^{(n)}=(1-n)^{-1}\log{\rm Tr}\,\rho_A^n$ which measure the
entanglement of an interval A of length $\ell$ with the rest of
an infinite one-dimensional system, in the case when this is
described by a conformal field theory of central charge $c$. These
can be due to bulk irrelevant operators of scaling dimension
$x>2$, in which case the leading corrections are of the expected
form $\ell^{-2(x-2)}$ for values of $n$ close to 1.  However for
$n>x/(x-2)$ corrections of the form $\ell^{2-x-x/n}$ and
$\ell^{-2x/n}$ arise and dominate the conventional terms. We also
point out that the last type of corrections can also occur with
$x$ less than 2. They arise from \em relevant \em operators
induced by the conical space-time singularities necessary to
describe the reduced density matrix. These agree with recent
analytic and numerical results for quantum spin chains. We also
compute the effect of marginally irrelevant bulk operators, which
give a correction $O((\log \ell)^{-2})$, with a universal
amplitude. We present analogous results for the case when the
interval lies at the end of a semi-infinite system.

\end{abstract}

\end{titlepage}

\subsection*{1. Introduction} Recently there has been much interest in
characterising bipartite quantum entanglement of pure states in
extended systems near a quantum critical point in terms of the
R\'enyi entropies\cite{Renyi}. For a given division of the Hilbert
space into a part A and its complement B, these are defined as
$$
S_A^{(n)}=(1-n)^{-1}\log{\rm Tr}\,\rho_A^n\,,
$$
where $\rho_A={\rm Tr}_B\,\rho$ is the reduced density matrix of
the subsystem A, and $\rho=|\Psi\rangle\langle\Psi|$ is the
density matrix of the whole system in a pure state $|\Psi\rangle$.
Knowledge of the $S_A^{(n)}$ for different $n$ characterizes the
full spectrum of non-zero eigenvalues of $\rho_A$ (see e.g.
\cite{cl-08}), and gives more information about the entanglement
than the widely studied von Neumann entropy $S_A^{(1)}$. It also
gives a fundamental insight into understanding the convergence and
scaling of algorithms based on matrix product states \cite{mps}.

In \cite{Holzhey,cc1,cc-rev} it was shown that for a one-dimensional
critical system whose scaling limit is described by a conformal
field theory (CFT) of conformal anomaly number (central charge)
$c$, in the case where A is an interval of length $\ell$ embedded
in an infinite system, the asymptotic behaviour of the R\'enyi
entropies is given by
\begin{equation}\label{Renyi:asymp}
S_A^{(n)}
\simeq \frac{c}6(1+n^{-1})\log \ell+O(1)\,.
\end{equation}
This result has by now been verified analytically and numerically
(see e.g.
\cite{Qspin,jk-04,km-04,zbfs-06,dmcf-06,ij-08,lk-08,ncc-08,afc-09,gl-rev,gt-10},
but this list is far from being exhaustive) for a large number of
examples of quantum spin chains whose scaling limit is believed to
be described by CFT. It gives, in fact, one of the most accurate
ways of measuring the conformal anomaly number. However, this
asymptotic result is often obscured by large (sometimes
oscillating) corrections to scaling \cite{recent,lsca-06,song}
whose origin has, so far, not been clarified in the context of
quantum field theory.

In any real system with an ultraviolet cut-off such as a lattice,
even when it is tuned to the critical point there will in general
be operators in the hamiltonian or action which ensure that the
continuum field theory results are only asymptotic on distance
scales much larger than the cut-off. It is important to understand
the form of the corrections to the asymptotic results in order to
make accurate fits to analytic and numerical finite-lattice data.
Renormalization group (RG) theory shows that these are ordinarily
due to irrelevant operators, with scaling dimension $x>d$ ($d$
being the dimension of space-time), which are allowed in the
effective action even at the critical point. These are generally
responsible for finite-size corrections in integer powers of
$L^{-(x-d)}$ \cite{FSS}, where $L$ is a characteristic length
scale of the system. Since field theory classifies these
irrelevant operators and allows the computation of their scaling
dimensions, it therefore is capable of quantifying the form of
corrections to scaling, if not the values of the non-universal
prefactors. Moreover, as $x\to2+$ these corrections become more
important, and, at $x=2$, the so-called marginal case, take the
form of logarithms rather than powers. The theoretical advantage
is that the amplitudes are then often universal and calculable.

In this paper we analyse the corrections to the R\'enyi entropies
due to irrelevant and marginally irrelevant bulk operators in the
hamiltonian. For the case of an interval of length $\ell$ in an
infinite system, we find the expected terms $O(\ell^{-2(x-2)})$.
For the case of an interval at the end of a semi-infinite system
there can also be terms $O(\ell^{-(x-2)})$. However we also find
unusual terms which are $O(\ell^{2-x-x/n})$ and $O(\ell^{-2x/n})$,
in which the exponent depends on $n$. For $n>n_c(x)=x/(x-2)$ these
dominate the conventional corrections.

While the appearance of $n$-dependent exponents for an irrelevant
bulk operator is perhaps surprising, in recent analytic and
numerical studies of several spin chains \cite{recent} even more
dramatic corrections of the form $\ell^{-2x/n}$, with values of
$x<2$, have been reported. We argue that these may also be
understood from the field theory under the assumption that the
conical singularities of the Riemann surfaces on which the path
integral must be evaluated in fact locally break the criticality
of the system, thus allowing the appearance of operators at these
points, which, in the bulk, would be \em relevant \em with scaling
dimension $x<2$. Such operators do not drive the system away from
bulk criticality because they are localised near points in
space-time. It is an old result of CFT
\cite{Cardy1984c}\footnote{In this reference, only corners on the
boundary were in fact considered. However, the analysis extends
trivially to the case of conical singularities in the bulk.} that
such a bulk operator near a conical singularity of degree $n$ in
fact has its scaling dimension modified to $x/n$. Thus we predict
that such operators should give rise to corrections of the form
$\ell^{-2x/n}$ in the case of an interval in an infinite system
and $\ell^{-x/n}$ in the semi-infinite case.

The marginal case $x=2$ turns out to be even more subtle and
difficult. However, we are able to show that the corrections take
a universal form in which $c$ in (\ref{Renyi:asymp}) is replaced
by
\begin{equation}\label{ceffR}
 c-\frac1{b^2(\log \ell)^3}+O\big((\log \ell)^{-4}\big)\,,
\end{equation}
where $b$ is a universal (and, for many CFTs, known) operator
product expansion (OPE) coefficient. Note that, for this leading
correction, the $n$-dependence is the same as that of the leading
term. As we discuss below, this is a consequence of
Zamolodchikov's $c$-theorem\cite{Zam}.

\subsection*{2. Field theory for corrections to scaling}

As was argued in Refs.~\cite{Holzhey,cc1,ccd-07}, for a
one-dimensional quantum system on a lattice $\cal L$ the
expression ${\rm Tr}\,\rho_A^n$ is given by the ratio $Z_n/Z_1^n$,
where $Z_n$ is the partition function defined by the path integral
on $n$ copies of ${\cal L}\otimes\mbox{imaginary time
$\tau\in\{-\infty,+\infty\}$}$, sewn together along $\tau=0$ so
that in the interval $B\subset{\cal L}$ the $j$th sheet with
$\tau>0$ is connected to the $j$th sheet with $\tau<0$, while in
the interval A the $j$th sheet is connected to the $(j+1)$th
sheet, cyclically. In the continuum limit, where the lattice is
replaced by the real line, this defines an $n$-sheeted Riemann
surface ${\cal R}_n$ with conical singularities at the ends of the
interval A, but in what follows later it is important to realise
that the identification also holds on a spatial lattice. It is
also convenient to consider the logarithm $-(F_n-nF_1)$ of the
above ratio of partition functions, which is then proportional to
the R\'enyi entropy $S_A^{(n)}$.

In analysing such a lattice system using continuum field theory
methods, the leading universal effects of the lattice are usually
assumed to be taken into account by imposing a short-distance
cut-off $\epsilon$ on the continuum theory: for example by
restricting the asymptotic expressions for correlation functions
to be valid only for separations greater than $\epsilon$. While
this might appear to be a rather crude approximation to the effect
of a lattice, since we are looking only for the universal form of
corrections to scaling and not their precise amplitudes, this is
in fact adequate. The response of the free energy $F$ to a change
in $\epsilon$ is in general given in the field theory by the
integrated trace of the stress tensor\cite{CardyPeschel,CardyLH}:
\begin{equation}\label{inttrace}
-\epsilon\frac{\partial F}{\partial\epsilon}=
\frac1{2\pi}\int\langle\Theta(z)\rangle d^2z\,.
\end{equation}
In a CFT in flat space-time, $\langle\Theta(z)\rangle=0$, implying
that $F$ is scale-invariant. (This is in fact strictly correct
only after subtracting off the non-universal bulk free energy,
which in our case automatically cancels in the combination
$F_n-nF_1$.) However, in Refs.~\cite{CardyPeschel,CardyLH} it was
shown that this is no longer true at conical singularities: in
fact each one contributes a term $(c/12)(n-n^{-1})$ to the right
hand side of (\ref{inttrace}). Integrating up and using the fact
that $F_n-nF_1$ can depend only on the ratio $\ell/\epsilon$ then
gives the result (\ref{Renyi:asymp}), which was derived in
slightly different ways in Refs.~\cite{Holzhey} and \cite{cc1}.

Let us now consider the effect of a bulk irrelevant operator in
the hamiltonian of action. In the field theory this is equivalent
to perturbing the CFT by an operator $\Phi(z)$ of scaling
dimension $x>2$, so the action is
\begin{equation}\label{pCFT}
S=S^*+\lambda\int\Phi(z)d^2z\,,
\end{equation}
where $S^*$ is the CFT action,  and $\lambda$ is a coupling
constant. It has the dimensional form $g/\epsilon^{2-x}$, where
$g$ is dimensionless. The change in the dimensionless free energy
is then formally given by the perturbative series
\begin{equation}\label{PT}
-\delta F_n=\sum_{N=1}^\infty\frac{(-\lambda)^N}{N!}\int_{{\cal
R}_n}\cdots\int_{{\cal R}_n}
\langle\Phi(z_1)\ldots\Phi(z_N)\rangle_{{\cal R}_n}\,d^2z_1\ldots
d^2z_N\,,
\end{equation}
of integrals of connected correlation functions of the CFT over
${\cal R}_n$.

In the case when A is the interval $(0,\ell)$ in an infinite
system, ${\cal R}_n$ may be conformally mapped to the punctured
complex plane ${\mathbb C}'={\mathbb C}\setminus\{0\}$
by\cite{cc1}
$$
\zeta=\left(\frac z{z-\ell}\right)^{1/n}:\qquad z=\ell f(\zeta)\equiv
\ell\frac{\zeta^n}{\zeta^n-1}\,.
$$
This maps the ends of the interval to $\zeta=0$ and $\infty$. The
correlation functions in the two geometries are related by
$$
\langle\Phi(z_1)\ldots\Phi(z_N)\rangle_{{\cal
R}_n}=\prod_{j=1}^N|\ell f'(\zeta_j)|^{-x}
\langle\Phi(\zeta_1)\ldots\Phi(\zeta_N)\rangle_{{\mathbb C}'}\,.
$$
Since $\langle\Phi\rangle_{{\mathbb C}'}=0$, the $N=1$ term in
(\ref{PT}) is absent in this case. The second order term may be
transformed to an integral over the $\zeta$-plane:
\begin{eqnarray}\label{Fn2}
\delta F_n^{(2)}&=&-\ffrac12g^2(\ell/\epsilon)^{4-2x}\int_{\mathbb
C'}\int_{\mathbb C'}\frac{|f'(\zeta_1)|^{2-x}|f'(\zeta_2)|^{2-x}}
{|\zeta_1-\zeta_2|^{2x}}d^2\zeta_1d^2\zeta_2
\\&=&-\ffrac12g^2(n\ell/\epsilon)^{4-2x}\int_{\mathbb
C'}\int_{\mathbb C'}\frac{|\zeta_1 \zeta_2|^{(2-x)(n-1)}}
{{|\zeta_1^n-1|}^{4-2x}  {|\zeta_2^n-1|}^{4-2x}
{|\zeta_1-\zeta_2|}^{2x}}d^2\zeta_1d^2\zeta_2 \,.\label{Fn3}
\end{eqnarray}
Note that this integral makes sense also for non-integer $n$,
although we derived it only for integer $n$. We may then consider
$n$ as an arbitrary real parameter. The integral has now several
potential sources of UV divergence which should be regulated by a
cutoff that in the $z$- plane is $O(\epsilon)$. These potential
divergences give rise to a further powers of $\epsilon$ in $\delta
F_n^{(2)}$ and, by scaling, to further powers in $\ell$ in the
corrections to scaling.

To elucidate the mechanism let us start from the case when $n-1$
is small and positive. The integral in (\ref{Fn3}) then converges
for $x<1$. For larger values of $x$, a divergence comes from the
region $\zeta_1\to\zeta_2$ and it should be regularised with a
cut-off $|z_1-z_2|<\epsilon$. The leading divergence in the
integral is $O(\epsilon^{2-2x})$, leading to a dependence in
$\delta F^{(2)}_n$ proportional to $\epsilon ^{-2}{\rm Area}({\cal
R}_n)$. This is a contribution to the non-universal bulk free
energy, which cancels in the combination $F_n-nF_1$. This
subtraction also cancels the apparent singularities at
$\zeta^n=1$, which correspond to $|z|\to\infty$ on ${\cal R}_n$.
If the leading divergence is subtracted off from (\ref{Fn3}), the
remainder is analytic at $x=1$. In fact, the finite part is then
given by the analytic continuation of (\ref{Fn3}) around its pole
at $x=1$. This is then finite all the way up to $x=3$, and the
$\ell$ dependence comes from the explicit prefactor, that is
$\ell^{-2(x-2)}$, the standard power law of FSS in the RG
framework. Note that although the amplitudes of these terms are
non-universal, depending on $g$, their ratios are universal, and
in principle calculable by evaluating the analytic continuation of
the integral. We also note that for an interval near the end of a
semi-infinite system, depending on the boundary conditions
$\langle\Phi\rangle$ may be non-vanishing, in which case the
leading correction will be $O(\ell^{2-x})$.

However, for larger values of $n$ and $x>2$, (\ref{Fn3}) may also
exhibit divergences as $\zeta_j\to 0$ or $\infty$. In the original
coordinates, these occur as $z_1$ or $z_2$ approach one of the
branch points. These are genuine divergences due to the conical
singularities and are not present in the bulk. Close to (say)
$z=0$ each integral behaves like $\int|z|^{((1/n)-1)x}d^2z$, which
diverges if $n>n_c(x)=x/(x-2)$. In this case the integral must be
further regulated with a cutoff $|z|<\epsilon$. This leads to an
further multiplicative factor $\propto\epsilon^{2-x+(x/n)}$ which,
once again, by scaling leads to corrections to scaling in the
R\'enyi entropy of the form
$\ell^{4-2x-(2-x+x/n)}=\ell^{2-x-x/n}$. For $n>n_c$ these
divergences are further enhanced when $z_1$ and $z_2$ are close to
different branch points, leading to a further factor of
$\epsilon^{2-x+x/n}$, and a consequent $\ell$-dependence of the
form $\ell^{4-2x-2(2-x+x/n)}=\ell^{-2x/n}$. (The singularities
when $z_1$ and $z_2$ approach the same branch point are removed by
the bulk free energy subtraction.)

To summarise, the presence of bulk irrelevant operators with $x>2$
leads to corrections to scaling of the form $\ell^{4-2x}$,
$\ell^{2-x-x/n}$ and $\ell^{-2x/n}$. In general, all these will be
present, but for $n<n_c=x/(x-2)$ the first dominates, while for
$n>n_c$ it is the last. For $n\approx n_c$ we expect them all to
play a role, with multiplicative logarithmic factors when $n=n_c$.

The appearance of these terms is similar to what happens in the
case of a boundary. In that case the bulk operator $\Phi(y)$ at a
distance $y$ from the boundary behaves like \cite{DD}
$y^{-x+x_b}\Phi_b$, where $\Phi_b$ is a boundary operator with
scaling dimension $x_b$. This gives rise, on integration, to a
term $\lambda^2\epsilon^{1-x+x_b}$ in the free energy per unit
length of the boundary which is singular if $x_b<x-1$, and hence
to a term $L^{x-x_b}\,L^{4-2x}$ in the total free energy of a
finite system of size $L$. This boundary contribution can actually
overwhelm the normal bulk term. Analogously, close to a conical
singularity, $\Phi(z)\sim |z|^{-x+x/n}\Phi^{(n)}(0)$, where
$\Phi^{(n)}$ is localised at the tip of the cone, and has scaling
dimension $x/n$. The difference is that in the boundary case $x_b$
is usually larger than the bulk scaling scaling dimension $x$,
while at a conical singularity the scaling dimension $(x/n)$
becomes arbitrarily small for large $n$.

\subsection*{3. Relevant operators at conical singularities}

We now argue that corrections to scaling of the form
$\ell^{-2x/n}$ in the R\'enyi entropy can arise not only by the
mechanism discussed above due to irrelevant bulk operators with
$x>2$, but in a different way which also gives corrections of this
form but with $x<2$. This is because a lattice model which is
critical can nevertheless generate operators localised at the
conical singularities with scaling dimension $(x/n)$ but with
$x<2$, that is, operators which, if they appeared in the bulk
hamiltonian, would be relevant and therefore drive the system away
from criticality. This most easily seen if we discretise time as
well as space. Consider, for example, a 2d classical model on a
square lattice. The anisotropic limit of this in general gives
rise to a 1d quantum hamiltonian. Computing ${\rm Tr}\,\rho_A^n$
for this quantum model corresponds, as before, to considering the
2d lattice model on an $n$-sheeted surface with branch points of
degree $n$. On the lattice, the details of this depend on exactly
how the degrees of freedom are divided between A and B. In most
models like quantum spin chains, the degrees of freedom are on the
lattice sites, so the division is along a spatial bond. In the
time-discretised picture, the branch point can be considered to
lie halfway along a spatial bond, or in the middle of a plaquette.
In either case, it is clear that degrees of freedom close to the
branch points have an enhanced (if $n>1$) number of neighbours. If
we maintain the same bond interactions as in the bulk, there will
therefore be an effective local coupling to the energy density, or
to other operators which do not break the internal symmetries of
the system. For example, in the case of the nearest-neighbour
Ising model, if we put the branch point in the centre of a
plaquette, each order variable (Ising spin) still has exactly 4
nearest neighbours, but the dual spin, which sits on top of the
branch point, will have $4n$ nearest neighbours, thus locally
breaking the self-duality of the model and, locally, driving it
away from criticality. In this case we would therefore expect a
coupling to the energy density, with $x=1$. In the field theory,
such an operator close to a conical singularity of degree $n$ has
its scaling dimension modified to $x/n$ \cite{Cardy1984c},
leading, in this case, to corrections of the form $\ell^{-2/n}$.

The conclusion is that the correct form of the field theory action
on the $n$-sheeted surface should be
$$
S=S_{\rm CFT}+\sum_j\lambda_j\int_{{\cal
R}_n}\Phi_jd^2z+\sum_P\sum_k\lambda_k\Phi^{(n)}_k(P)\,,
$$
where the second term takes account of bulk irrelevant operators
$\Phi_j$ with $x_j>2$, already discussed, and the third term is a
sum over the branch points $P$ of localised operators
$\Phi^{(n)}_k(P)$ with scaling dimension $x_k/n$, with all
possible values of $x_k$ allowed by symmetry, including those with
$x_k<2$. However, in the perturbative expansion in powers of the
$\lambda_k$, each $\Phi^{(n)}_k(P)$ at a given branch point should
appear at most once (otherwise we can use the OPE to write higher
powers in terms of other localised operators). In the case of an
infinite system, since $\langle\Phi^{(n)}(P)\rangle=0$, we
therefore expect the leading correction to be $O(\ell^{-2x_k/n})$.
This should be the case no matter how many branch points $\geq2$
there are. For an interval at the end of a semi-infinite system,
depending on the boundary conditions it may be that
$\langle\Phi^{(n)}(P)\rangle\not=0$, in which case the leading
correction will be $O(\ell^{-x_k/n})$.

\subsection*{4. The marginal case and the $c$-theorem}

Next we turn to the marginal case $x\to2+$, which is technically
more challenging. The integral in (\ref{Fn2}) is in general very
difficult to manipulate into a form in which the necessary
analytic continuation around the pole at $x=1$ can be performed.
It is easier to extract the limit $x\to2$ by making the
subtraction explicitly, and this we now describe. It is, however,
quite subtle, as the cut-off and subtraction must be performed in
the $z$-plane. Imposing a cut-off $|z_1-z_2|>\epsilon$, in the
$\zeta$-plane we have
\begin{equation}\label{deltaFn2}
\delta F_n=-\ffrac12g^2(\ell/\epsilon)^{4-2x}\int
\int_{|f(\zeta_1)-f(\zeta_2)|>(\epsilon/\ell)}\frac{|f'(\zeta_1)|^{2-x}|f'(\zeta_2)|^{2-x}}
{|\zeta_1-\zeta_2|^{2x}}d^2\zeta_1d^2\zeta_2\,.
\end{equation}
The subtraction is
\begin{eqnarray*}
-\ffrac12g^2(\ell/\epsilon)^{4-2x}\int_{{\cal R}_n}
\frac{2\pi\epsilon^{2-2x}}{2-2x}d^2z_1&=&-\ffrac12g^2(\ell/\epsilon)^{4-2x}\int
|f'(\zeta_1)|^2\frac{2\pi\epsilon^{2-2x}}{2-2x}d^2\zeta_1\\&=&
-\ffrac12g^2(\ell/\epsilon)^{4-2x}\int
\int_{|\zeta_2-\zeta_1|>(\epsilon/\ell |f'(\zeta_1)|)}
\frac{|f'(\zeta_1)|^{4-2x}}{|\zeta_2-\zeta_1|^{2x}}d^2\zeta_1d^2\zeta_2\,.
\end{eqnarray*}
It can be shown that, up to terms which vanish as $x\to2$, the
cut-off in the last integral can be replaced by
$|f(\zeta_1)-f(\zeta_2)|>(\epsilon/\ell)$. Thus the second order
contribution to the subtracted free energy is
\begin{eqnarray*}
&&\delta F_n-n\delta F_1 =\\&&-\ffrac12g^2(\ell/\epsilon)^{4-2x}\int
\int_{|f(\zeta_1)-f(\zeta_2)|>(\epsilon/\ell)}
\frac{|f'(\zeta_1)|^{2-x}|f'(\zeta_2)|^{2-x}-|f'(\zeta_1)|^{4-2x}}
{|\zeta_1-\zeta_2|^{2x}}d^2\zeta_1d^2\zeta_2\,.
\end{eqnarray*}
Symmetrising between $\zeta_1$ and $\zeta_2$ we then find
\begin{equation}\label{gg}
\delta F^{(2)}_n-n\delta
F^{(2)}_1=\ffrac14g^2(\ell/\epsilon)^{4-2x}\int
\int_{|f(\zeta_1)-f(\zeta_2)|>(\epsilon/\ell)}
\frac{\Big(|f'(\zeta_1)|^{2-x}-|f'(\zeta_2)|^{2-x}\Big)^2}
{|\zeta_1-\zeta_2|^{2x}}d^2\zeta_1d^2\zeta_2\,.
\end{equation}
The integral is now finite as $\epsilon\to0$ and the cutoff can be
removed, at least for $x$ close to 2. Also, notice that everywhere
$f'(\zeta_1)$ and $f'(\zeta_2)$ are regular, the integrand
vanishes as $x\to2$. However this is not the case close to the
conical singularities. For example, near $\zeta=0$,
$|f'(\zeta)|^{2-x}\sim |\zeta|^{(n-1)(2-x)}$, and the limits
$\zeta\to0$ and $x\to2$ do not commute. However we would obtain
the same limit as $x\to2$ in the integral if we restricted the
integration region to, say,
$(|\zeta_1|<\rho|,|\zeta_2|<\rho)\cup(|\zeta_1|>\rho^{-1}|,|\zeta_2|>\rho^{-1})$
for any $0<\rho<1$. In particular, we can take $\rho$ arbitrarily
small, in which case we can accurately approximate $f'(\zeta)$ by
its asymptotic forms $n\zeta^{n-1}$ and $n\zeta^{-n-1}$, up to
terms which vanish as $x\to2$. The consequence\footnote{This also
shows that the argument also holds in presence of a boundary when
the form $|\zeta_1-\zeta_2|^{-2x}$ of the two-point function holds
only at short distances.} is that, as $x\to2$, we can replace the
integral in (\ref{deltaFn2}) by the analytic continuation to
$x\sim 2$ of
$$
2\int\int\frac{|n\zeta_1^{n-1}|^{2-x}|n\zeta_2^{n-1}|^{2-x}}{|\zeta_1-\zeta_2|^{2x}}d^2\zeta_1d^2\zeta_2\,.
$$
This, apart from the $\ell^{4-2x}$ prefactor, is precisely twice what
we would obtain for a single branch point. This integral can be
evaluated explicitly. Rescaling $\zeta_2=w\zeta_1$ we have
\begin{equation}\label{zetaI}
2n^{4-2x}\,I(n,x)\,\int |\zeta_1|^{-2-2n(x-2)}d^2\zeta_1\,,
\end{equation}
where\footnote{This follows using the representation
$|w|^{2a}=\Gamma(-a)^{-1}\int_0^\infty u^{-1-a} e^{-uw^*w}du$ for
each factor and first performing the gaussian integral over $w$.
The result can then be reduced to a beta-function integral.}
\begin{eqnarray*}
I(n,x)&=&\int|w|^{(n-1)(2-x)}|w-1|^{-2x}d^2w\\
&=&\pi\frac{\Gamma\big(1+(n+1)(x-2)/2\big)\Gamma\big(1-(n-1)(x-2)/2\big)\Gamma(1-x)}
{\Gamma\big(-(n+1)(x-2)/2\big)\Gamma\big((n-1)(x-2)/2\big)\Gamma(x)}\,.
\end{eqnarray*}
Although the equivalence of this result to (\ref{gg}) is valid
only as $x\to2$, we note that it does exhibit the required poles
at $x=1$ and $x=2n/(n-1)$, correspond to the short-distance
divergences already discussed above. However, as $x\to2$ we have
$$
I(n,x)\sim -(\pi/4)(n^2-1)(x-2)+O\big((x-2)^2\big)\,.
$$
The integral over $\zeta_1$ in (\ref{zetaI}), gives, after
imposing a short-distance cutoff $|\zeta|>\epsilon^{1/n}$, a
factor
$$
\frac{2\pi\big(\epsilon^{1/n}\big)^{2n(x-2)}}{2n(x-2)}\sim
\frac\pi{n(x-2)}+O(1)\,.
$$
Putting all these factors together we then find that, for
$x\approx2$
\begin{equation}\label{FF}
F_n-nF_1=-\frac c6(n-n^{-1})\log(\ell/\epsilon)+
g^2(n-n^{-1})(\ell/\epsilon)^{4-2x}\left(\frac{\pi^2}4+O(x-2)\right)+O(g^3)\,.
\end{equation}
For $x=2$ this gives an (apparently) uninteresting constant
contribution to the R\'enyi entropy, and it is therefore necessary
to go the next order, involving an integral over the 3-point
function of the form
$$
\ffrac16bg^3(\ell/\epsilon)^{6-3x}\int_{\mathbb
c}\frac{|f'(\zeta_1)|^{2-x}|f'(\zeta_1)|^{2-x}|f'(\zeta_1)|^{2-x}}
{|\zeta_2-\zeta_3|^x|\zeta_3-\zeta_1|^x|\zeta_1-\zeta_2|^x}
d^2\zeta_1d^2\zeta_2d^2\zeta_3\,,
$$
where $b$ is the universal coefficient in the OPE
$$
\Phi(\zeta)\cdot\Phi(0)=|\zeta|^{-2x}\big(1+b|\zeta|^x\Phi(0)+\cdots\big)\,.
$$
Once again, it can be shown that, after subtracting the
short-distance singularities as $\zeta_j\to\zeta_k$, the measure
is concentrated on the conical singularities as $x\to2$, so the
result for the interval is twice that found by replacing
$f(\zeta)$ by $\zeta^n$. The integral can then be performed
explicitly. (A similar computation was carried in
Refs.~\cite{Cardylog,CardyLudwig} for the corrections to the free
energy of a cylinder, which corresponds to the limit $n\to0$ of
the present calculation.) However, it turns out that the
coefficient of the $O(g^3)$ term in (\ref{FF}) is determined from
the $O(g^2)$ term by Zamolodchikov's $c$-theorem\cite{Zam}. To use
this, however, it is more correct to consider the logarithmic
derivative of the free energy with respect to the cut-off, which,
from (\ref{inttrace}), is proportional to the integral of the
trace $\langle\Theta\rangle$: we see from (\ref{FF}) that this
takes the form $-(c_{\rm eff}(g)/6)(n-n^{-1})$ where
\begin{equation}\label{ceff2}
c_{\rm eff}(g)=c-3\pi^2(2-x)g^2+O(g^3)\,.
\end{equation}
The $c$-theorem\cite{Zam} states that there exists a function
$C(g)$ which decreases along RG flows and is stationary at fixed
points where it equals the conformal anomaly number of the
corresponding CFT. Zamolodchikov's analysis also implies that
$C'(g)\propto\big(1+O(g^2)\big)\beta(g)$, where
$\beta(g)=-\epsilon(\partial g/\partial\epsilon)_{\lambda_R}$,
keeping the renormalised coupling $\lambda_R$ fixed. For a
perturbed CFT, the first two terms in $\beta(g)$ are
universal\cite{Zam,Cardybook}:
$$
-\beta(g)=(2-x)g-\pi bg^2+O(g^3)\,.
$$
Thus, up to and including terms $O(g^3)$, all candidates for an
interpolating function $C(g)$ must agree, and in particular
$c_{\rm eff}'(g)\propto\beta(g)$ to this order. This fixes the
coefficient of the $O(g^3)$ term in (\ref{ceff2}) to be
$2\pi^3b$.\footnote{This result is in fact universal in all
physical manifestations of perturbative corrections to $c$, for
example the conformal anomaly $\langle\Theta\rangle=-cR/12$ in a
curved background of scalar curvature $R$ \cite{Cardyprep}.}

However, the result in (\ref{ceff2}) disguises the fact that in
the neglected higher order terms logarithmic dependences on $\ell$
appear, corresponding to poles at $x=2$. These can, however, be
absorbed by ``RG-improving'' the expansion, that is, replacing $g$
by $g(\ell)$ where
$$
\ell dg(\ell)/d\ell =-\beta\big(g(\ell))\,.
$$
If $x>2$, that is, the perturbation is slightly irrelevant, then
asymptotically $g(\ell)\propto \ell^{-(x-2)}$ and $c_{\rm
eff}(\ell)\to c$, with a power-law correction, consistent with our
earlier analysis. If $x<2$, that is the perturbation is relevant,
then, depending on the sign of $g/b$, either $g(\ell)\to g^*$ and
$c_{\rm eff}(\ell)\to c_{\rm
new}=c-(2-x)^3/b^2+O((2-x)^4)$,\cite{CardyLudwig} or $g(\ell)$
grows beyond the range of this perturbative treatment.

The interesting case is when $x=2$, that is, the perturbation is
marginal. Then if $g/b<0$, the perturbation is marginally relevant
and $g(\ell)$ again grows, but if $g/b>0$ it is marginally irrelevant
and $g(\ell)$ flows to zero, albeit logarithmically slowly. In fact
$$
g(\ell)=\frac g{1+\pi bg\log(\ell/\epsilon)}\sim\frac1{\pi
b\log(\ell/\epsilon)}\,.
$$
Substituting this into (\ref{ceff2}) we then see that
\begin{equation}\label{log}
c_{\rm
eff}(\ell)=c+\frac2{b^2(\log(\ell/\epsilon))^3}+O\big((\log(\ell/\epsilon))^{-4}\big)\,.
\end{equation}
If we now integrate with respect to $\epsilon$ to find the free
energy $F_n-nF_1$, we find the result in (\ref{ceffR}). Note that
this asymptotic value is reached from below, in apparent
contradiction to Zamolodchikov's $c$-theorem. However, the correct
definition of $C(g)$ in this case is through the logarithmic
derivative of the entanglement entropy \cite{CasiniHuerta}, which
is ultraviolet finite at the fixed points, and in this quantity
$c$ is approached from above. We also remark that, for fitting any
finite $\ell$ data, it is better to replace the asymptotic
$\big(\log(\ell/\epsilon)\big)^{-3}$ in (\ref{log}) by
$g(\ell)^3$.

\subsection*{5. Comparison with other studies and discussions}

We have shown that the block entanglement in an infinite 1d system
(measured by  the R\'enyi entropies $S_A^{(n)}$) generically
displays standard corrections to scaling of the form
$\ell^{-2(x-2)}$ with $x>2$ and {\it unusual} ones of the form
$\ell^{-2x/n}$ (and also the `combination' $\ell^{2-x-x/n}$, which
however is never the leading one). We call these corrections {\it
unusual} because of the explicit $n$ dependence of the exponent, a
property that seems to contrast with RG finite size scaling theory
\cite{FSS}. Clearly there is no contradiction: we showed that
these terms arise from the conical singularities needed to
describe $S_A^{(n)}$ in a path integral formulation. The existence
of such geometry-dependent exponents was in fact first noticed a
long time ago in Ref.~\cite{Cardyedge}. The most surprising effect
here is that the unusual correction $\ell^{-2x/n}$ can be present
also for relevant operators with $0<x<2$, occasioned by a {\it
local} breaking of scale invariance at the conical singularity.
Such effects have probably not be seen in earlier studies of
corner critical behaviour \cite{corner} because they focussed on
effective values of $n<1$ where the exponent of these corrections
is larger.

These unusual effects clearly need direct confirmation from
lattice computations. Large corrections to scaling have been
observed in several numerical studies quoted before, but a
quantitative analysis has become available only recently
\cite{recent}. It has been shown analytically for the Ising and XX
universality class that corrections to the scaling are of the form
$\ell^{-2/n}$ \cite{recent}, which agrees with our general
formula when $x=1$. For the Ising model, $x=1$ corresponds to
the energy density operator that indeed we argued to be generated
by the conical singularity. For anisotropic Heisenberg chains, the
corrections to the scaling have been found numerically to be of
the form $\ell^{-2K/n}$, where $K$ is the Luttinger liquid
exponent, i.e. the most relevant present in the continuum theory.
Again this perfectly agrees with our result. Similar unusual (i.e.
$n$ dependent) corrections have been also found in other
entanglement measures \cite{ar-10,k-09}, but their quantitative
understanding in the framework of quantum field theory requires
further investigation.

In the case of systems with boundaries, we showed that these
unusual corrections are of the form $\ell^{-x/n}$. Evidence of
such power laws has been reported for $n=1$ \cite{lsca-06,song}
both for XX and Heisenberg chains, but a quantitative study for
general $n$ is still lacking. Preliminary results show that
our results are correct \cite{recent,ce-prep}.

However, our theory of the origin of the $\ell^{-2x/n}$
corrections with $x<2$ (and $\ell^{-x/n}$ in the semi-infinite
case) has a definite \em prediction\em, so far untested in
analytic and numerical studies of particular systems: if the
origin of these terms is indeed in the local deviation from
criticality near the conical singularities, then a modification of
the lattice action close to the singularities should have the
effect of changing the \em amplitudes \em of these corrections. In
fact, by tuning the local couplings to a particular value (which
may however depend on $n$) it should be possible to eliminate
these correction terms altogether. It should be stressed that this
modification needs to be done locally in space-time, not simply by
modifying the interaction strengths in the hamiltonian near the
ends of the interval. This could be carried out by starting with
the ground state $|0\rangle$ of the full hamiltonian $\hat H$, and
evolving this with the operator $\exp(-\tau{\hat H}')$, where
${\hat H}'$ is $\hat H$ with modified interactions in a region
$O(\epsilon\ll\ell)$ around the ends of the interval. Here
$\tau\sim\epsilon/v$, with $v$ being the coefficient in the
quasiparticle dispersion relation $\omega\sim v|k|$. The
prediction is that the leading behaviour of the R\'enyi entropies
measured in the modified state is that same as that of the ground
state, but that the amplitudes of the $O(\ell^{-2x/n})$ correction
terms should be different.

It is also worth commenting on the other correction
$\ell^{2-x-x/n}$, that is never the most relevant one and could be
obscured by the others in numerical analysis. For $n\sim n_c$ its
effect should be more important, but an accurate quantitative
analysis is difficult because the various corrections get too
close to each other. However in Ref. \cite{recent} it has been
noticed that for $n$ close to 1 and for $\Delta<-0.5$ (the
anisotropy parameter of the Heisenberg chain), the single
correction $\ell^{-2K/n}$ does not describe numerical data
accurately. This is a confirmation of the presence of other
corrections to the scaling that could be of the form
$\ell^{2-x-x/n}$ (but could also be of very different origin).

We stress that these corrections in some lattice systems (those
described by Luttinger liquid theory) show a strong oscillating
behaviour \cite{recent,lsca-06,lk-08,r-2}. While the physical
origin of this effect is due to strong tendency to
antiferromagnetic order \cite{lsca-06},  a general proof of this
universal form from the continuum Luttinger liquid field theory is
still lacking.

Marginal perturbations deserve  separate discussion, because of
the logarithmic corrections to the scaling. It is a well-known
(and obvious) fact that log-corrections are hard to detect in
numerical studies, and in fact, up to now no direct evidence for
them has been still provided. However, for the isotropic
Heisenberg antiferromagnet (that has a marginal operator)  the
relations found in \cite{lsca-06} between entanglement entropy
($n=1$) and energy of the ground-state (known to have similar kind
of log-corrections) is an indirect evidence of the correctness of
our prediction (cf. Eq. (\ref{log})).

Finally we want to comment on the case of more than one interval,
corresponding to more than two branch points in the $n$-sheeted
Riemann surface. In this case, already the leading term (the
analogue of Eq. (\ref{Renyi:asymp})) is rather involved
\cite{cg-08,fps-08,cct-09,ip-09}, but it is calculable for integer
$n$ for the simplest CFT's \cite{fps-08,cct-09,atc-09}. Even in
this case, the large corrections to the scaling prevent a direct
simple analysis \cite{fps-08,cct-09} and the check of the
complicated asymptotic forms. The exact knowledge of the
correction to the scaling exponents, would greatly simplify the
analysis. Up to now, it has been shown numerically for the Ising
model, with periodic boundary conditions, and for $n=2$ that the
exponent is $1/2$ \cite{atc-09}, compatible with a form such as
$2x/n$ ($x=1/2$ and $n=2$). Exploiting the exact solvability of
the model, an accurate analysis \cite{fc-10} shows also for other
low values of $n=3,4$ corrections compatible with $1/n$. In the
Ising model, the non-local fermion operator has dimension $1/2$
and could be the origin of such scaling. In fact, to build the
reduced density of the spin degrees of freedom in the Ising model,
one should introduce the non local Jordan-Wigner string between
the two intervals \cite{atc-09,ip-09,fc-10}, that plays the rule
of the Ising fermion. This does not enter in the single interval
reduced density matrix that is the same for spin and fermion
degrees of freedom. As a confirmation of this behaviour, we
mention that for fermions in the Ising model we have the same
correction as for the single interval $2/n$ \cite{fc-10}, because
the Jordan-Wigner string is not present. For the XX model, the
same analysis \cite{fc-10} shows unambiguously that the exponent
is exactly $2/n$. Thus all evidences confirm that also for two
intervals the correction to the scaling exponent is of the form
$2x/n$, even if the example of the Ising model shows that some
care is needed to fix the appropriate value of $x$.

{\it Acknowledgments}. We thank F.~Essler for very helpful
discussions. This work was supported by the EPSRC under grant
EP/D050952/1 (JC).

\end{document}